\documentclass[prl,reprint]{revtex4-1}
\pdfoutput=1
\usepackage{graphicx}
\usepackage{bm}% bold math
\usepackage{amsmath}
\usepackage{hyperref}% add hypertext capabilities
\begin{document}

\title{Evolution of the $7/2$ fractional quantum Hall state in two-subband systems}
\date{today}

\author{Yang Liu}
\affiliation{Department of Electrical Engineering,
Princeton University, Princeton, New Jersey 08544}
\author{J.\ Shabani}
\affiliation{Department of Electrical Engineering,
Princeton University, Princeton, New Jersey 08544}
\author{D.\ Kamburov}
\affiliation{Department of Electrical Engineering,
Princeton University, Princeton, New Jersey 08544}
\author{M.\ Shayegan}
\affiliation{Department of Electrical Engineering,
Princeton University, Princeton, New Jersey 08544}
\author{L.N.\ Pfeiffer}
\affiliation{Department of Electrical Engineering,
Princeton University, Princeton, New Jersey 08544}
\author{K.W.\ West}
\affiliation{Department of Electrical Engineering,
Princeton University, Princeton, New Jersey 08544}
\author{K.W.\ Baldwin}
\affiliation{Department of Electrical Engineering,
Princeton University, Princeton, New Jersey 08544}

\date{\today}

\begin{abstract}

  We report the evolution of the fractional quantum Hall state (FQHS)
  at even-denominator filling factor $\nu=7/2$ in wide GaAs quantum
  wells in which electrons occupy two electric subbands.  The data
  reveal subtle and distinct evolutions as a function of density,
  magnetic field tilt-angle, or symmetry of the charge distribution.
  When the charge distribution is strongly asymmetric, there is a
  remarkable persistence of a resistance minimum near $\nu=7/2$ when
  two Landau levels belonging to the two subbands cross at the Fermi
  energy. The field position of this minimum tracks the 5/2 filling of
  the symmetric subband, suggesting a pinning of the crossing levels
  and a developing 5/2 FQHS in the symmetric subband even when the
  antisymmetric level is partially filled.

\end{abstract}

%\pacs{}

\maketitle

The fractional quantum Hall states (FQHSs) at the even-denominator
Landau level (LL) filling factors \cite{Willett.PRL.1987} have
recently come into the limelight thanks to the theoretical prediction
that these states might be non-Abelian \cite{Moore.Nuc.Phy.1991} and
be useful for topological quantum computing
\cite{Nayak.Rev.Mod.Phys.2008}. This expectation has spawned a flurry
of investigations, both experimental \cite{ Dean.PRL.2008,
  Choi.PRB.2008, Nuebler.PRB.2010, Shabani.PRL.2010, Kumar.PRL.2010,
  Xia.PRL.2010, Pan.PRL.2011} and theoretical \cite{Rezayi.PRL.2000,
  Peterson.PRL.2008,*Peterson.PRB.2008, Wojs.PRL.2010}, into the
origin and stability of the even-denominator states. Much of the
attention has been focused on the $\nu=5/2$ FQHS which is observed in
very low disorder two-dimensional electron systems (2DESs) when the
Fermi energy ($E_F$) lies in the spin-up, excited-state ($N=1$), LL of
the ground-state (symmetric, S) electric subband of the 2DES, namely
in the S1$\uparrow$ level. Here we examine the stability of the FQHS
at $\nu=7/2$, another even-denominator FQHS, typically observed when
$E_F$ is in the S1$\downarrow$ level
(Fig. 1(a))\cite{Dean.PRL.2008,Shabani.PRL.2010}. The $\nu=7/2$ FQHS,
being related to the 5/2 state through particle-hole symmetry, is also
theoretically expected to be non-Abelian. Our study, motivated by
theoretical proposals that the even-denominator FQHSs might be favored
in 2DESs with "thick" wavefunctions
\cite{Rezayi.PRL.2000,Peterson.PRL.2008,*Peterson.PRB.2008,Wojs.PRL.2010},
is focused on electrons confined to wide GaAs quantum wells (QWs). In
a realistic, experimentally achievable wide QW, however, the electrons
at $\nu=7/2$ can occupy the second (antisymmetric, A) electric subband
when the subband energy spacing ($\Delta$) is comparable to the
cyclotron energy $\hbar\omega_c$ (Figs. 1(b-d)). Here we
experimentally probe the stability of the $\nu=7/2$ FQHS in wide QW
samples with tunable density in the vicinity of the crossings (at
$E_F$) between the S1 and the A0 LLs.

Our samples were grown by molecular beam epitaxy, and each consist of
a wide GaAs QW bounded on each side by undoped
Al$_{0.24}$Ga$_{0.76}$As spacer layers and Si $\delta$-doped
layers. We report here data, taken at $T\simeq30$ mK, for three
samples with QW widths of $W =$ 37, 42, and 55 nm. The QW width and
electron density ($n$) of each sample were designed so that its
$\Delta$ is close to $\hbar\omega_c$ at the magnetic field position of
$\nu=7/2$. This enables us to make the S1 and the A0 LLs cross at
$E_F$ by tuning $n$ or the charge distribution asymmetry, which we
achieve by applying back- and front-gate biases
\cite{Shabani.PRL.2010,Suen.PRL.1994,Shabani.PRL.2009,Liu.cond.mat.2011}. For
each $n$, we measure the occupied subband electron densities $n_S$ and
$n_A$ from the Fourier transforms of the low-field ($B\le 0.5$ T)
Shubnikov-de Haas oscillations \cite{Suen.PRL.1994,Shabani.PRL.2009},
and determine $\Delta=(\pi\hbar^2/m^{*})(n_S-n_A)$, where $m^{*} =
0.067m_e$ is the GaAs electron effective mass. At a fixed total
density, $\Delta$ is smallest when the charge distribution is
"balanced" (symmetric) and it increases as the QW is imbalanced. Our
measured $\Delta$ agree well with the results of calculations that
solve the Poisson and Schroedinger equations to obtain the potential
energy and the charge distribution self-consistently (see, e.g.,
Figs. 1(a,d)).

\begin{figure}
\includegraphics[width=0.48\textwidth]{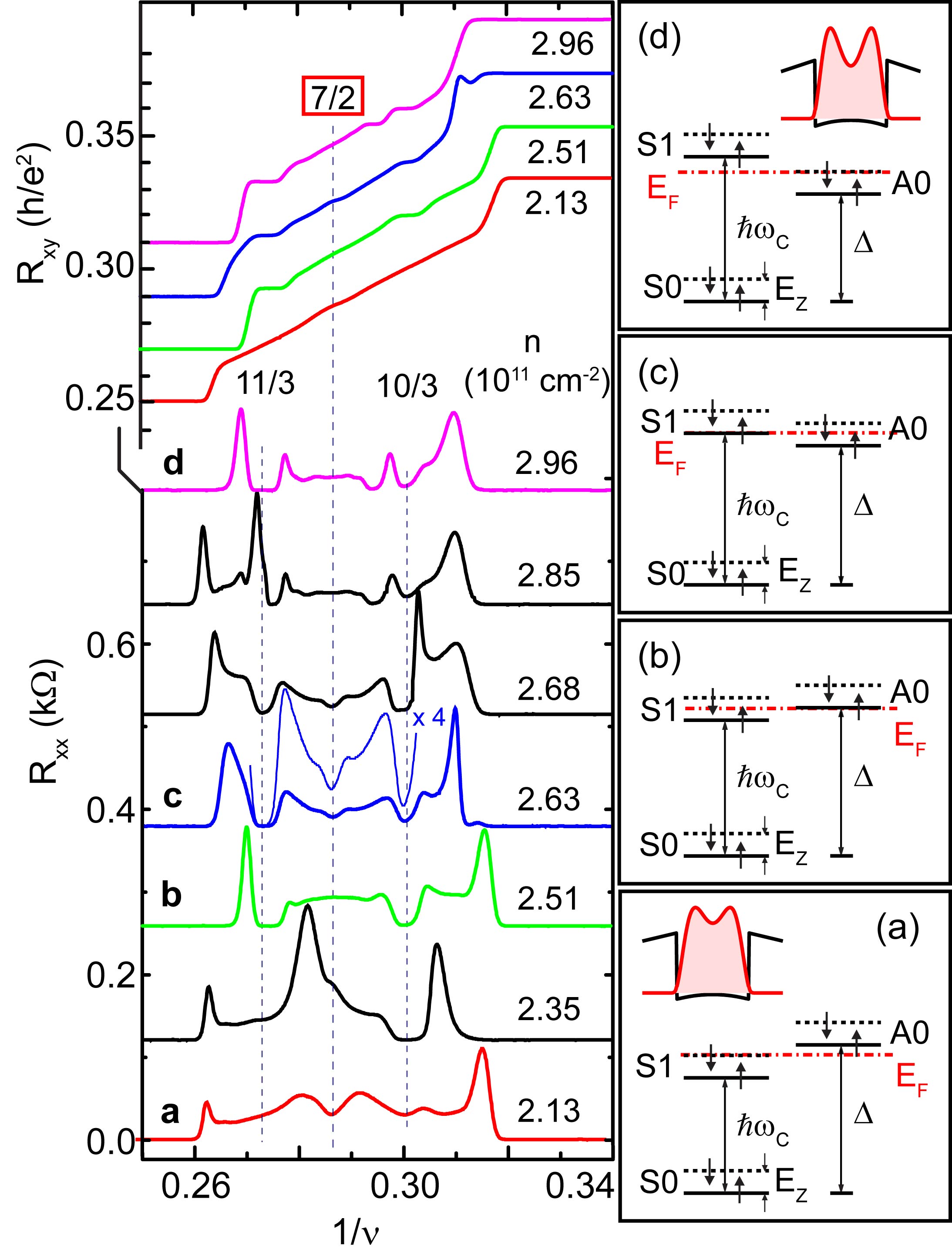}
\caption{\label{fig:balanced} (color online) Left panel: Waterfall
  plot of $R_{xx}$ and $R_{xy}$ traces at different densities for a
  42-nm-wide GaAs QW. (a-d) LL diagrams at $\nu=7/2$ for different
  densities. Self-consistently calculated charge distributions are
  shown in the insets to (a) and (d) for $n=2.13$ and $2.96\times
  10^{11}$ cm$^{-2}$.}
\end{figure}

Figure 1 shows a series of longitudinal ($R_{xx}$) and Hall ($R_{xy}$)
resistance traces in the range $3 < \nu < 4$ for a 42 nm-wide QW
sample, taken at different $n$ from $2.13$ to $2.96\times 10^{11} $
cm$^{-2}$ while keeping the total charge distribution balanced. As $n$
is increased in this range, $\Delta$ decreases from 64 to 54 K while
$\hbar\omega_c$ at $\nu=7/2$ increases from 50 K to 70 K, so we expect
crossings between the S1 and A0 levels, as illustrated in
Figs. 1(a-d). These crossings manifest themselves in a remarkable
evolution of the FQHSs as seen in Fig. 1. At the lowest $n$, which
corresponds to the LL diagram shown in Fig. 1(a), $R_{xx}$ shows a
reasonably deep minimum at $\nu$ = 7/2, accompanied by a clear
inflection point in $R_{xy}$ at $7/2(h/e^2)$, and a weak minimum near
$\nu=10/3$. These features are characteristic of the FQHSs observed in
high-quality, standard (single-subband) GaAs 2DESs, when $E_F$ lies in
the S1$\downarrow$ LL \cite{Dean.PRL.2008,Shabani.PRL.2010}. As $n$ is
raised, we observe an $R_{xx}$ spike near $\nu=7/2$, signaling a
crossing of S1$\downarrow$ and A0$\uparrow$. At $n=2.51\times 10^{11}
$ cm$^{-2}$, these levels have crossed, and $E_F$ is now in
A0$\uparrow$ (Fig. 1(b)). There is no longer a minimum at $\nu=7/2$
and instead, there are very strong minima at $\nu=10/3$ and
11/3. Further increasing $n$ causes a crossing of S1$\uparrow$ and
A0$\uparrow$ and, at $n=2.63\times 10^{11} $ cm$^{-2}$, $E_F$ at
$\nu=7/2$ lies in S1$\uparrow$ (Fig. 1(c)). Here the $R_{xx}$ minimum
and $R_{xx}$ inflection point at $\nu=7/2$ reappear, signaling the
return of a FQHS. As we increase $n$ even further, S1$\uparrow$ and
A0$\downarrow$ cross and, at $n=2.96\times 10^{11} $ cm$^{-2}$, when
$E_F$ at $\nu=7/2$ lies in A0$\downarrow$, there is again no $\nu=7/2$
minimum but there are strong FQHSs at $\nu=10/3$ and 11/3.

The above observations provide clear and direct evidence that the
even-denominator $\nu=7/2$ FQHS is stable when $E_F$ is in an excited
($N=1$) LL but not when $E_F$ lies in a ground-state ($N=0)$ LL
\cite{Shabani.PRL.2010}. Examining traces taken at numerous other $n$,
not shown in Fig. 1 for lack of space, reveal that the appearances and
disappearances of the $\nu=7/2$ FQHS are sharp, similar to the
behavior of the $5/2$ FQHS at a LL crossing
\cite{Liu.cond.mat.2011B}. It is noteworthy that when the two crossing
levels have $antiparallel$ spins, a "spike" in $R_{xx}$ at the
crossing completely destroys the FQHS at $\nu=7/2$ and nearby
fillings. At the crossing of two levels with $parallel$ spins, on the
other hand, there is no $R_{xx}$ spike. These behaviors are
reminiscent of easy-axis and easy-plane ferromagnetism for the
antiparallel- and parallel-spin crossings, respectively
\cite{Muraki.PRL.2001, Liu.cond.mat.2011}. 
%Also, the $\nu=7/2$ FQHS
%appears to be stronger when $E_F$ lies in the S1$\uparrow$ level
%compared to the S1$\downarrow$ level; this is evident from the deeper
%$R_{xx}$ minimum and better formed $R_{xy}$ plateau at $n=2.63\times
%10^{11}$ cm$^{-2}$ compared to $n=2.13\times 10^{11}$ cm$^{-2}$.

\begin{figure}
\includegraphics[width=0.4\textwidth]{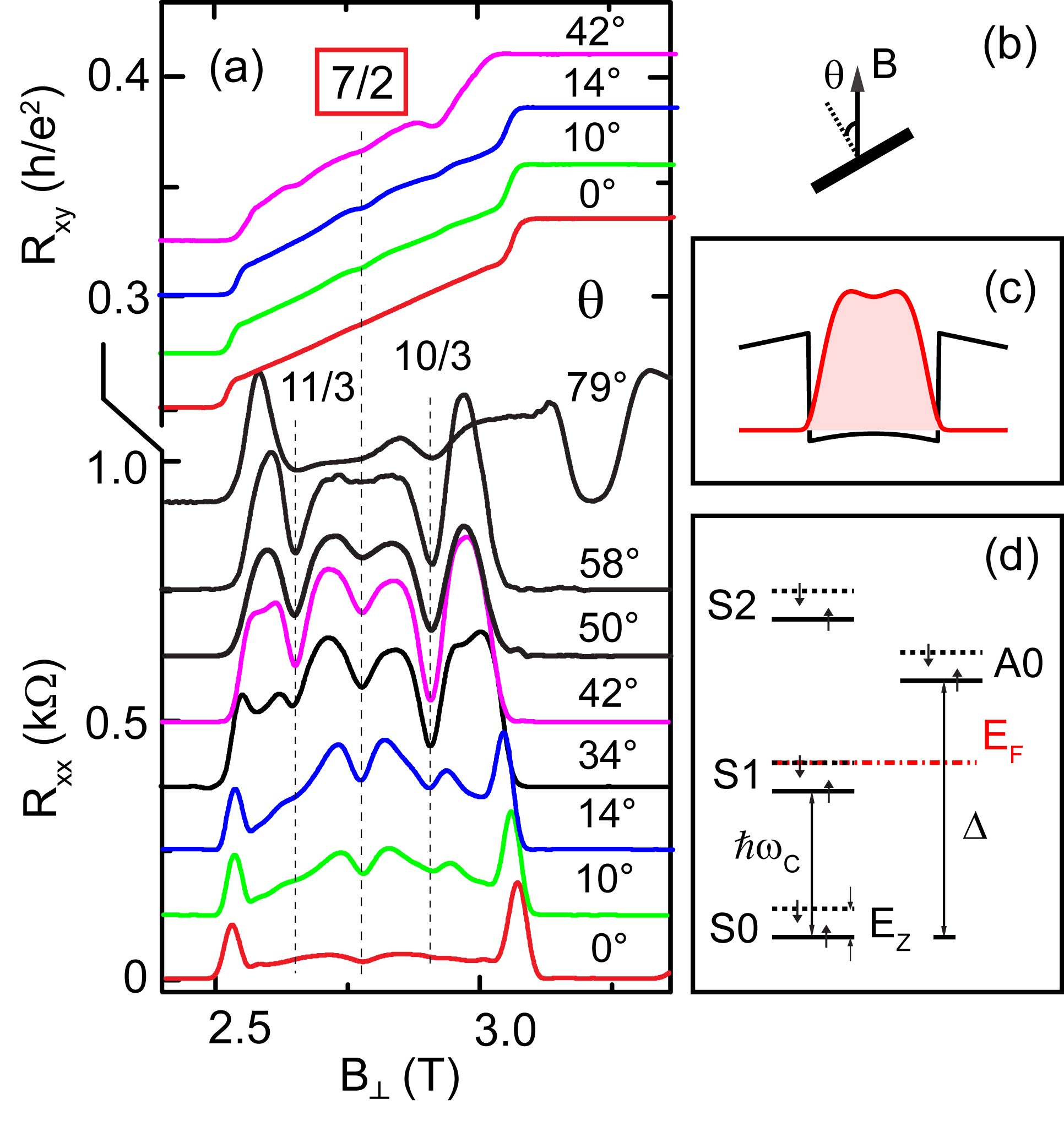}
\caption{\label{fig:tilted} (color online) (a) $R_{xx}$ and $R_{xy}$
  traces for a 37-nm-wide GaAs QW at $n=2.34\times 10^{11}$ cm$^{-2}$
  at different tilt angles $\theta$ as depicted in (b). (c) Charge
  distribution calculated self-consistently at $B=0$. (d) LL diagram
  at $\theta=0$ at $\nu=7/2$.}
\end{figure}

Next, we examine the evolution of the $\nu=7/2$ FQHS in the presence
of a parallel magnetic field component $B_{||}$, introduced by tilting
the sample so that its normal makes an angle $\theta$ with the total
field direction (Fig. 2(b)). Figure 2(a) captures this evolution for
electrons confined to a symmetric, 37-nm-wide QW \footnote{Application
  of a $B_{||}$ component leads to an anisotropic state at
  intermediate tilt angles. Traces shown in Fig. 2 were taken along
  the hard-axis; data along the easy-axis show a qualitatively similar
  behavior.}. This QW is narrower so that, at $n=2.34\times 10^{11}$
cm$^{-2}$, its $\Delta$ (= 82 K) is well above $\hbar\omega_c$ (= 55
K). The $\theta=0$ trace then corresponds to $E_F$ lying in
S1$\downarrow$, as shown in Fig. 2(d). As $\theta$ is increased, we
observe only a gradual change in the strength of the $\nu=7/2$ FQHS,
until it disappears at large $\theta \gtrsim$ 55$^\circ$. This is not
surprising since, in a two-subband system like ours, we expect a
severe mixing of the LLs of the two subbands with increasing $\theta$
\cite{Kumada.PRB.2008} rather than sharp LL crossings as manifested in
Fig. 1 data.

We highlight three noteworthy features of Fig. 2 data. First, the
$\nu=7/2$ $R_{xx}$ minimum persists up to relatively large
$\theta$ (up to $50^\circ$), and it even appears that the $R_{xy}$
plateau is better developed at finite $\theta$ (up to
$\theta=42^\circ$) compared to $\theta=0$, suggesting a strengthening of
the 7/2 FQHS at intermediate angles. Second, deep $R_{xx}$
minima develop with increasing $\theta$ at $\nu=10/3$ and 11/3,
suggesting the development of reasonably strong FQHSs at these
fillings. This is consistent with the results of Xia $et$ $al.$ who
report a similar strengthening of the 7/3 and 8/3 states - the
equivalent FQHSs flanking the $\nu=5/2$ state in the S1$\uparrow$
level - when a wide QW sample is tilted in field
\cite{Xia.PRL.2010}. Third, the large magnitude of $B_{||}$ at the highest
angles appears to greatly suppress $\Delta$ \cite{Hu.PRB.1992},
rendering the electron system essentially into a bilayer system. This
is evidenced by the dramatic decrease in the strength of the $\nu=3$
QHS and the disappearance of the $\nu=11/3$ $R_{xx}$
minimum at $\theta=79^\circ$; note that a FQHS should not exist at $\nu=11/3$ in a
bilayer system with two isolated 2DESs as it would correspond to 11/6
filling in each layer.

\begin{figure}
\includegraphics[width=.48\textwidth]{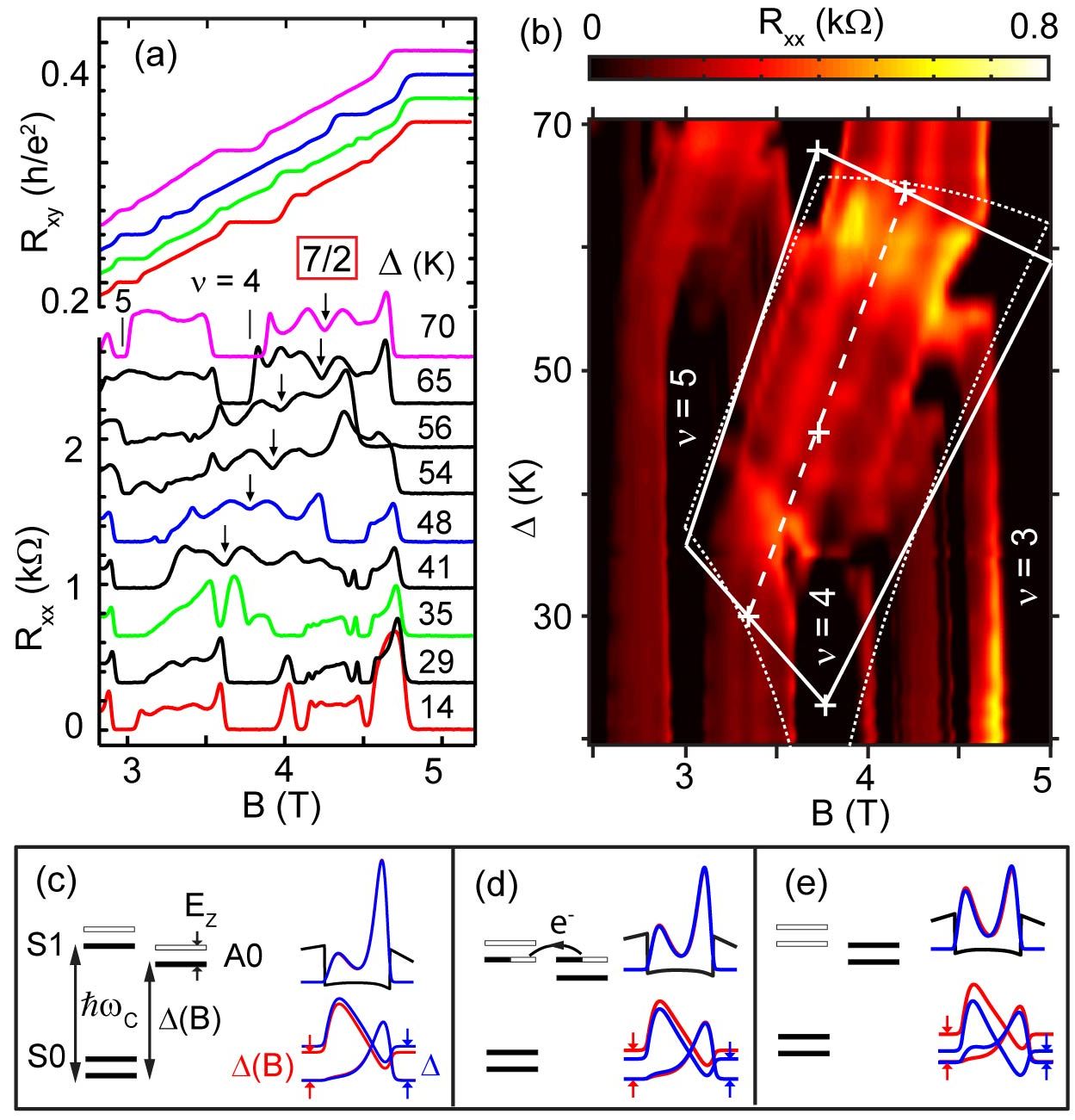}
\caption{ \label{fig:imbalancing} (color online) (a) $R_{xx}$ and
  $R_{xy}$ traces for a 55-nm-wide GaAs QW at a fixed $n=3.62\times
  10^{11}$ cm$^{-2}$, as the charge distribution is made increasingly
  asymmetric. Values of $\Delta$, measured from low-$B$ Shubnikov-de
  Haas oscillations, are indicated for each trace. Vertical arrows
  mark the positions of observed anomalous $R_{xx}$ minima. (b) A
  color-scale plot of data in (a). Solid and dotted lines are the
  calculated boundary within which the S1$\uparrow$ and A0$\uparrow$
  levels are pinned together at $E_F$. The dashed line represents the
  values of $B$ at which, according to the calculations, the
  S1$\uparrow$ level is half-filled; it tracks the positions of the
  observed $R_{xx}$ minima marked by vertical arrows in (a). (c-e)
  Schematic LL diagrams (left) and charge distributions and
  wavefunctions (right), self-consistently calculated at $B=0$ (blue)
  and at $\nu=4$ (red). In (c-e), the filling factor of the
  S1$\uparrow$ level equals 1, 0.5, and 0, respectively. In each
  panel, the calculated wavefunctions are shifted vertically according
  to the calculated values of $\Delta$ and $\Delta(B)$.}
\end{figure}

We now focus on data taken on a 55-nm-wide QW where we keep the total
$n$ fixed and change the charge distribution symmetry by applying
back- and front-gate biases with opposite polarity. In Fig. 3(a) we
show a set of $R_{xx}$ traces, each taken at a different amount of
asymmetry. The measured $\Delta$ is indicated for each trace and
ranges from 14 K for the symmetric charge distribution to 70 K for a
highly asymmetric distribution. In Fig. 3(b) we present a color-scale
plot of $R_{xx}$ with $B$ and $\Delta$ as $x$ and $y$ axes, based on
an interpolation of Fig. 3(a) data and many other traces taken at
different values of $\Delta$. When the charge distribution is
symmetric or nearly symmetric in this QW, $\Delta$ is much smaller
than $\hbar\omega_c$ (= 85 K at $\nu=7/2$) so that the LL diagram is
qualitatively the one shown in Fig. 1(d). Consistent with this LL
diagram, we observe a very strong $\nu=4$ QHS. Also, since $E_F$ lies
in the A0$\downarrow$ level at $\nu=7/2$, there is no $\nu=7/2$ FQHS
and instead we observe strong FQHSs at $\nu=10/3$ and 11/3. As
$\Delta$ is increased, we expect a crossing of S1$\uparrow$ and
A0$\downarrow$, leading to a destruction of the $\nu=4$ QHS at the
crossing. This is indeed seen in Figs. 3(a) and (b). What is striking,
however, is that the $\nu=4$ $R_{xx}$ minimum disappears over a very
large range of $\Delta$, between 35 and 62 K. Even more remarkable are
several anomalous $R_{xx}$ minima in this range of $\Delta$ in the
filling range $3<\nu<5$, particularly those marked by arrows in
Fig. 3(a). These minima resemble what is observed in the top trace but
are seen at lower fields.

These features betray a pinning together, at $E_F$, of the partially
occupied S1$\uparrow$ and A0$\downarrow$ levels, and a change transfer
between them, in a finite range of $B$ and gate bias. As pointed out
in Ref. \cite{Trott.PRB.1989}, when only a small number of quantized
LLs belonging to two different subbands are occupied, the distribution
of electrons between these levels does not necessarily match the $B=0$
subband densities. This leads to a mismatch between the total electron
charge density distributions at $B=0$ and high $B$. The pinning and
the inter-LL charge transfer help bring these distributions closer to
each other \cite{Trott.PRB.1989, Davies.PRB.1996,
  *Solovyev.PRB.2009}. To demonstrate such a pinning quantitatively
and determine the boundary inside which the S1$\uparrow$ and
A0$\downarrow$ levels are pinned together, we performed
self-consistent calculations of the potential energy and charge
distribution at high $B$, similar to those described in
Ref. \cite{Trott.PRB.1989}. This boundary is marked by solid lines in
Fig. 3(b).

 % \footnote{A. G. Davies \emph{et al.},
 %  Phys. Rev. B {\bf 54}, R17331 (1996); V. V. Solovyev \emph{et al.},
 %  Phys. Rev. B {\bf 80}, 241310 (2009).}

First, we assume that one point at this boundary, corresponding to
A0$\downarrow$ having just moved above S1$\uparrow$ at $\nu=7/2$ so
that S1$\uparrow$ is half-filled and A0$\downarrow$ is empty, occurs
at $\Delta=65$ K. The rationale is that, for $\Delta=65$ K, the
$B=0$ subband density ratio $n_S/n_A=2.5$ is equal to the ratio of
fillings $\nu_S/\nu_A$ $(=2.5/1.0)$ for the S and A subbands at
$\nu=7/2$. Note in Fig. 3 that the trace taken at $\Delta=65$ K
indeed shows a strong minimum at total $\nu=7/2$ but traces taken at
lower $\Delta$ do not. This is consistent with our assumption that
$\Delta\simeq 65$ K marks the onset of full depopulation of
A0$\uparrow$. Now, since the onset occurs when
$\Delta=\hbar\omega_c-E_Z$, we conclude that $E_Z\simeq 21$ K,
implying an effective $g$-factor $\simeq$ 7.3, which is 17-fold
enhanced over the GaAs band value (0.44); this is similar to the
enhancements observed for electrons confined to similar wide QWs
\cite{Liu.cond.mat.2011}.

Next, we assume that in the region where the pinning occurs, $g$ has a
fixed value of 7.3 \footnote{The fact that the boundary we calculate
  based on $g=7.3$ matches the region where the pinning is observed
  validates our choice of this value. A full many-body calculation that
  takes exchange interaction and variations of $g$ with $B$ and
  wavefunction asymmetry into account might explain the experimental
  observations more accurately.}, and that the $in$-$field$
subband separation is given by $\Delta(B)=\hbar\omega_c-E_Z$. This
expression ensures that $\Delta(B)$ is fixed at a given $B$,
consistent with the pinning of the S1$\uparrow$ and A0$\downarrow$
levels. We then perform in-field self-consistent calculations for a
series different QW asymmetries. For each QW asymmetry, the in-field
charge distribution is given by:
\begin{equation}
  \rho(B)=e(eB/h)[\nu_S\cdot|\psi_S(B)|^2+\nu_A\cdot|\psi_A(B)|^2].
\end{equation}
Now, for the different points on the boundary, $\nu_S$ and $\nu_A$
have specific and well-defined values. For example, at $\nu=4$
($B=3.75$ T), for which we show the results of our self-consistent
calculations in Figs. 3(c) and (e), we have $\nu_S=3$ and $\nu_A=1$
for the upper boundary and $\nu_S=\nu_A=2$ for the lower
one. Focusing on the upper boundary, i.e., using $\nu_S=3$ and
$\nu_A=1$ in Eq. (1), among all the QW asymmetries for which we
perform the self-consistent calculations, one in particular has a
subband separation which is equal to $\Delta(B)$ (= 56 K for
$B=3.75$ T). This particular QW asymmetry gives the upper boundary at
$B=3.75$ T. We then calculate the zero-field subband separation for
this asymmetry, which turns out to be $\Delta=68$ K, and mark it in
Fig. 3(b) as the upper boundary for the pinning at $B=3.75$ T. For the
lower boundary at $B=3.75$ T, we repeat the above calculations using
$\nu_S=\nu_A=2$. The QW asymmetry that gives $\Delta(B)=56$ K yields
a zero-field $\Delta$ of 23 K which we mark in Fig. 3(b) as the
lower boundary at 3.75 T. The rest of the boundary in Fig. 3(b) is
determined in a similar fashion. For example, the upper boundary at
$\nu=7/2$ corresponds to ($\nu_S=2.5$, $\nu_A=1$) and the lower
boundary to ($\nu_S=2$, $\nu_A=1.5$).

It is clear that the calculated boundary marked by the solid lines in
Fig. 3(b) matches well the region (in $\Delta$ vs. $B$ plane) in which
we experimentally observe a disappearance of the $\nu=4$ $R_{xx}$
minimum and the appearance of $R_{xx}$ minima at anomalous
fillings. This matching is particularly remarkable, considering that
there are no adjustable parameters in our simulations, except for
using a single value (7.3) for the enhanced $g$-factor \cite{Note2}. In
Fig. 3(b) we also include a dashed line representing the values of $B$
at which, according to our calculations, the S1$\uparrow$ level is
exactly half-filled, i.e., $\nu_S=5/2$ and $\nu_A=(\nu-5/2)$. This
dashed line tracks the positions of the observed $R_{xx}$ minima
marked by vertical arrows in (a) very well, suggesting that these
minima indeed correspond to $\nu_S=5/2$. This is an astonishing
observation, as it implies that there is a developing FQHS at 5/2
filling of the symmetric subband even when a partially filled
A0$\downarrow$ level is pinned to the half-filled S1$\uparrow$ level
at $E_F$!

In Fig. 3(b) we also show a boundary, marked by dotted lines, which is
based on a simple, $analytic$ model. Note that the simulations
shown in Figs. 3(c-e) indicate that the in-field charge distributions,
calculated self-consistently at $\nu=4$, are nearly the same as the
$B=0$ distributions. In our simple model, we assume that the in-field
wavefunctions $\psi_S(B)$ and $\psi_A(B)$ are just linear combinations
of the $B=0$ wavefunctions $\psi_S(0)$ and $\psi_A(0)$. We then
set the total in-field charge distribution, given by Eq. (1), equal to
its $B=0$ value, $\rho(0)=en_S|\psi_S(0)|^2+en_A|\psi_A(0)|^2$, and find:
\begin{equation}
  \label{eq:relation}
  \Delta_0^2=(\nu_S-\nu_A)(eB/h)(\pi\hbar^2/m^{*})\Delta(B).
\end{equation}
For a given value of $B$ and therefore $\Delta(B)=\hbar\omega_c-E_Z$,
Eq. (2) gives the $B=0$ subband separation $\Delta_0$ which
corresponds to the onset of pinning/depinning of the relevant LLs. For
example, to find the upper boundary at $B=4$ T ($\nu=3.75$), we use
$\Delta(B)=60$ K, $\nu_A=1$ and $\nu_S=2.75$, and solve Eq. (2) to
find $\Delta_0=65$ K. To find the lower boundary at $B=4$ T, we use
$\nu_S=2$ and $\nu_A=1.75$ and find $\Delta_0=25$ K. As seen in
Fig. 3(b), the dotted line given by the simple, analytic expression
(2) matches the boundary determined from in-field self-consistent
calculations reasonably well except for the lower points where
$\nu_S=\nu_A\simeq 2$ leads to $\Delta_0\simeq 0$.

In summary our results reveal distinct metamorphoses of the
ground-state of two-suband 2DESs at and near $\nu=7/2$ as either the
field is tilted, or the density or the charge distribution symmetry
are varied. Most remarkably, we observe a developing FQHS when a
half-filled S1$\uparrow$ level is pinned to a partially-filled
A0$\downarrow$ level \footnote{J. Nuebler \emph{et al.} independently
  made a similar observation, that a FQHS exists when the S1$uparrow$
  level is half filled while the A0$uparrow$ level is partially
  occupied (unpublished).}.

% Near the completion of our work, we learned of a similar
%   observation, namely that a FQHS exists when $\nu_S=5/2$ while the
%   A0$uparrow$ level is partially occupied, by J. Nuebler $et$ $al.$
%   (unpublished).
% The most astonishing is perhaps the observation
% of a developing FQHS when the symmetric subband has a 5/2 filling
% while the A0$\downarrow$ is pinned to S0$\uparrow$ partially full
% .

\begin{acknowledgments}
  We acknowledge support through the DOE BES (DE-FG0200-ER45841) for
  measurements, and the Moore Foundation and the NSF (DMR-0904117 and
  MRSEC DMR-0819860) for sample fabrication and characterization. A
  portion of this work was performed at the National High Magnetic
  Field Laboratory, which is supported by the NSF, DOE, and the State
  of Florida.
\end{acknowledgments}

% \nocite{*}
\bibliography{paper_v2}

\end{document}